\documentclass[journal]{IEEEtran}
\DeclareUnicodeCharacter{202F}{\,}
\DeclareUnicodeCharacter{2009}{\,}
\DeclareUnicodeCharacter{2248}{\approx}

\usepackage{graphicx}
\usepackage{array}
\usepackage{booktabs}
\usepackage{tabularx}
\usepackage{threeparttable}
\usepackage{textcomp}
\usepackage{stfloats}
\usepackage{cite}
\usepackage{url}
\usepackage{orcidlink}
\usepackage{soul}
\usepackage{pifont}
\usepackage[dvipsnames]{xcolor}
\usepackage{microtype}
\def\BibTeX{{\rm B\kern-.05em{\sc i\kern-.025em b}\kern-.08em
    T\kern-.1667em\lower.7ex\hbox{E}\kern-.125emX}}

\begin{document}

\title{Quantum-Enhanced Learning Framework for Intelligent and AI-Native 6G Wireless Networks}

\author{Muhammad Bilal Akram Dastagir\,\orcidlink{0000-0003-2990-4604}, Rakesh Saini\,\orcidlink{0000-0002-4963-8465}, Omer Tariq\,\orcidlink{0000-0002-1771-6166},\\
Saif~Al-Kuwari\,\orcidlink{0000-0002-4402-7710}, Shahid Mumtaz\,\orcidlink{0000-0001-6364-6149}, Anwer Al-Dulaimi\,\orcidlink{0000-0001-6364-6149}, and Ahmed Farouk\,\orcidlink{0000-0001-8702-7342}
\thanks{Muhammad Bilal Akram Dastagir is with the Qatar Center for Quantum Computing, College of Science and Engineering, Hamad Bin Khalifa University, Doha, Qatar. E-mail: mdastagir@hbku.edu.qa}
\thanks{Rakesh Saini is with the Qatar Center for Quantum Computing, College of Science and Engineering, Hamad Bin Khalifa University, Doha, Qatar. E-mail: rasa68842@hbku.edu.qa}
\thanks{Omer Tariq is with the School of Computing, Korea Advanced Institute of Science and Technology, Daejeon, South Korea. E-mail: omertariq@kaist.ac.kr}
\thanks{Saif Al-Kuwari is with the Qatar Center for Quantum Computing, College of Science and Engineering, Hamad Bin Khalifa University, Doha, Qatar. E-mail: smalkuwari@hbku.edu.qa}
\thanks{Shahid Mumtaz is with Nottingham Trent University, Engineering Department, United Kingdom. E-mail: dr.shahid.mumtaz@ieee.org}
\thanks{Anwer Al-Dulaimi is with Veltris, Toronto, Canada. E-mail: anwer.aldulaimi@ieee.org}
\thanks{Ahmed Farouk is with the Qatar Center for Quantum Computing, College of Science and Engineering, Hamad Bin Khalifa University, Doha, Qatar, and the Department of Computer Science, Faculty of Computers and Artificial Intelligence, Hurghada University, Hurghada, Egypt. E-mail: ahmedfarouk@ieee.org}
}


\maketitle
\begin{abstract}
The recent convergence of 6G wireless systems and Tiny Machine Learning (TinyML) has driven the need for on-device intelligence in edge networks, where ultra-low latency, stringent energy budgets, and tight compute constraints demand novel architectures. Lightweight deep models efficiently extract local patterns but fail to capture global dependencies, while attention mechanisms do so at the expense of energy and computational cost. To bridge this gap, we introduce Quantumer, a hybrid TinyML--quantum framework that integrates multi-scale dilated convolutions and scaled dot-product attention within a lightweight transformer architecture, employing a two-stage transfer learning pipeline from Quantum Pre-Training (Quantumer-Q) to Classical Fine-Tuning (Quantumer-C). We also present QuantiblentLayer, a four-qubit variational circuit that maps compact traffic representations into measurement-based Hilbert-space features using trainable rotations and cyclic entangling operations. The circuit is used only during offline pre-training as a nonlinear embedding teacher and is removed before Quantumer-C deployment, leaving a fully classical inference model without runtime quantum execution. By transferring these quantum-assisted embeddings into an energy-efficient, lightweight transformer, Quantumer achieves strong detection performance with minimal compute and memory overhead on resource-constrained edge devices. The intrusion detection system (IDS) is used as a case study and evaluated on the Edge-IIoTset, TON IoT, and WUSTL-IIoT-2021 datasets. Quantumer-Q achieves competitive compact-model performance with 105.86K parameters, 0.4038 MB memory usage, 0.5525 MB model size, and 5.5646 MFLOPs; the INT8 Raspberry Pi 4 deployment obtains 16.8413 ms latency with a 0.6493 MB footprint. These results support training-time quantum-assisted representation learning for compact, classical, and edge-deployable IDS in 6G IIoT networks.
\end{abstract}


\begin{figure}[!t]
  \centering
  \includegraphics[width=\linewidth]{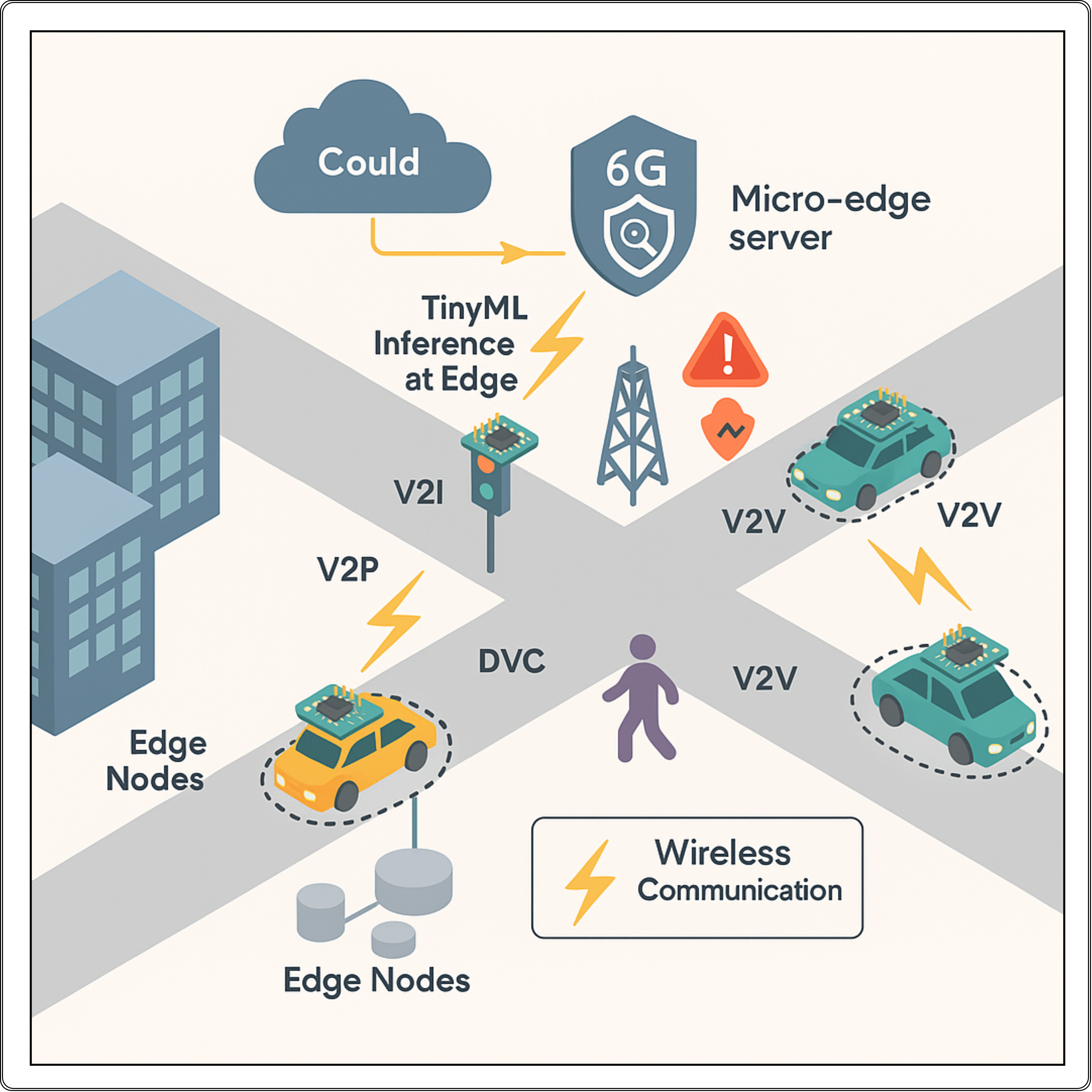}
  \caption{Conceptual overview of Quantumer: 6G-enabled edge nodes (vehicles, RSUs, sensors) perform on-device TinyML inference, while a 6G micro-edge server supports quantum-assisted intrusion detection for real-time and energy-aware network security.}
  \label{Fig:conceptIDS}
\end{figure}

\section{Introduction}
\label{sec:introduction}

The convergence of 6G wireless systems and small-scale machine learning (``small ML'') is redefining the role of intelligence in edge-driven networks. These next-generation systems must support adaptive real-time decision-making while operating under strict constraints on latency, energy consumption, and processing power. In this evolving setting (see Fig.~\ref{Fig:conceptIDS}), deploying learning models at the network edge becomes a key enabler of autonomous and resilient 6G applications~\cite{Saad2019}. From an industrial engineering perspective, embedding ML at the wireless edge presents a multi-dimensional challenge, requiring co-optimization of model design, system architecture, and wireless protocol behavior. Solutions must address changing traffic conditions, intermittent connectivity, hardware variability, and security requirements. Pioneering frameworks such as MCUNet show how joint algorithm--compiler--hardware design can enable inference on microcontrollers, yet the complexity of network intrusion detection in 6G remains insufficiently addressed~\cite{Shi2016,Serror2021}. Tiny Machine Learning (TinyML), though well established in academia, represents one segment of a larger ``small ML'' design space of resource-efficient frameworks. These operate in confined on-device or distributed collaborative modes in edge devices, IoT sensors, drones, wearables, and intelligent infrastructure, raising questions about model architecture, deployment strategy, cross-layer integration, communication cost, and task offloading in 6G wireless networks~\cite{Warden2019}.

\subsection{Limitations of Classical Transformers for TinyML-Edge Applications and Motivation}
Although the Transformer architecture is effective for modeling long-range dependencies through self-attention, it incurs quadratic computation and memory growth with sequence length, which can be unsuitable for TinyML edge deployments under strict 6G resource budgets~\cite{Vaswani2017}. Its parameter count and computational cost make on-device inference challenging for hardware with limited memory and low operating frequency.

\subsubsection{Research Gap: Toward Quantum-Enhanced Embeddings}
Recent lightweight transformer variants, including TSLANet's adaptive spectral blocks~\cite{TSLANet_ICML} and NanoMST's hardware-aware multiscale embeddings~\cite{NanoMST_11053971}, reduce parameters and computation, but remain fully classical and rely on restricted embedding strategies. The remaining gap is how to use a compact quantum feature map during training while keeping the deployed IDS model fully classical and small enough for embedded inference.

\subsubsection{Research Objective}
Our objective is to design an energy-efficient lightweight transformer enhanced with quantum Hilbert-space learning representation, which (i) addresses the computational and energy inefficiencies of classical transformers on resource-limited edge devices, (ii) enriches feature embeddings through nonlinear quantum mappings~\cite{Biamonte2017,Havlicek2019}, and (iii) supports compact inference within 6G network constraints. The objective is not to claim quantum speedup during inference. Instead, we examine whether a small variational circuit can act as a nonlinear embedding teacher during training, improve representation structure, and then be removed before deployment. Furthermore, we adapt the hybrid quantum-classical architecture fusion methodology, for which preliminary results show practical benefits in complex data classification~\cite{QCHybrid_DL_10716490}.

\subsection{Proposed Quantumer Architecture}
Quantumer follows a two-stage pipeline comprising Quantum Pre-Training (Quantumer-O) and Classical Fine-Tuning (Quantumer-C). In Quantumer-O, a multiscale 1D convolutional module and compact transformer blocks extract temporal features from packet sequences and project them into four rotation angles that control a four-qubit variational circuit. During training, the circuit produces four measurement-based outputs. These outputs guide supervised pre-training, whereas the quantum block is removed before deployment. The resulting learned interface is then fine-tuned in Quantumer-C using an energy-efficient transformer, enabling end-to-end TinyML inference.

\subsubsection{Advantages of Quantumer for TinyML-Edge 6G Applications}
By combining quantum-assisted training with a hardware-aware transformer backbone, Quantumer provides nonlinear feature extraction during pre-training and a lightweight classical inference path during deployment. This design separates representation learning from deployment costs: quantum computation is used only during training, while the final edge model remains classical, compact, and quantizable.

\subsubsection{Evaluation and Comparative Performance}
To validate Quantumer, we conduct experiments on three benchmark IDS datasets: Edge-IIoTset~\cite{Edge-IIoTset_IDS_Dataset}, TON IoT~\cite{TON_IoT_IDS_Dataset}, and WUSTL-IIoT-2021~\cite{WUSTL-IIOT-2021_IDS_Dataset}. The evaluation reports not only detection metrics, but also representation separability, model size, parameter count, memory usage, MFLOPs, GPU latency, quantized model behavior, and Raspberry Pi 4 inference latency. Quantumer-Q uses only 105.86K parameters and 5.5646 MFLOPs, while the INT8 Raspberry Pi 4 deployment obtains 16.8413 ms inference latency with a 0.6493 MB model footprint.

\subsection{Major Contributions}
The major contributions of this work are as follows:
\begin{enumerate}
  \item \textbf{Quantum Hilbert-Space Learning Representation:} We introduce a quantum-assisted embedding framework that projects classical network telemetry into a variational quantum circuit during training, enabling nonlinear feature shaping beyond purely classical embeddings.
  \item \textbf{Energy-Efficient Lightweight Transformer Architecture:} We design a compact transformer backbone tailored for TinyML-edge devices, reducing computational complexity and memory footprint while preserving the ability to capture longer-range traffic dependencies.
  \item \textbf{Training-Time QuantiblentLayer with Classical Deployment:} We use QuantiblentLayer as a four-qubit training-time embedding module and remove it during deployment so that the final Quantumer-C model remains fully classical and suitable for edge inference.
  \item \textbf{Deployment-Aware IDS Evaluation:} We evaluate Quantumer on Edge-IIoTset, TON IoT, and WUSTL-IIoT-2021 and report feature separability, confusion matrices, ROC behavior, model complexity, INT8 deployment behavior, and Raspberry Pi 4 latency.
\end{enumerate}

\begin{figure*}[!htbp]
  \centering
  \includegraphics[width=\textwidth]{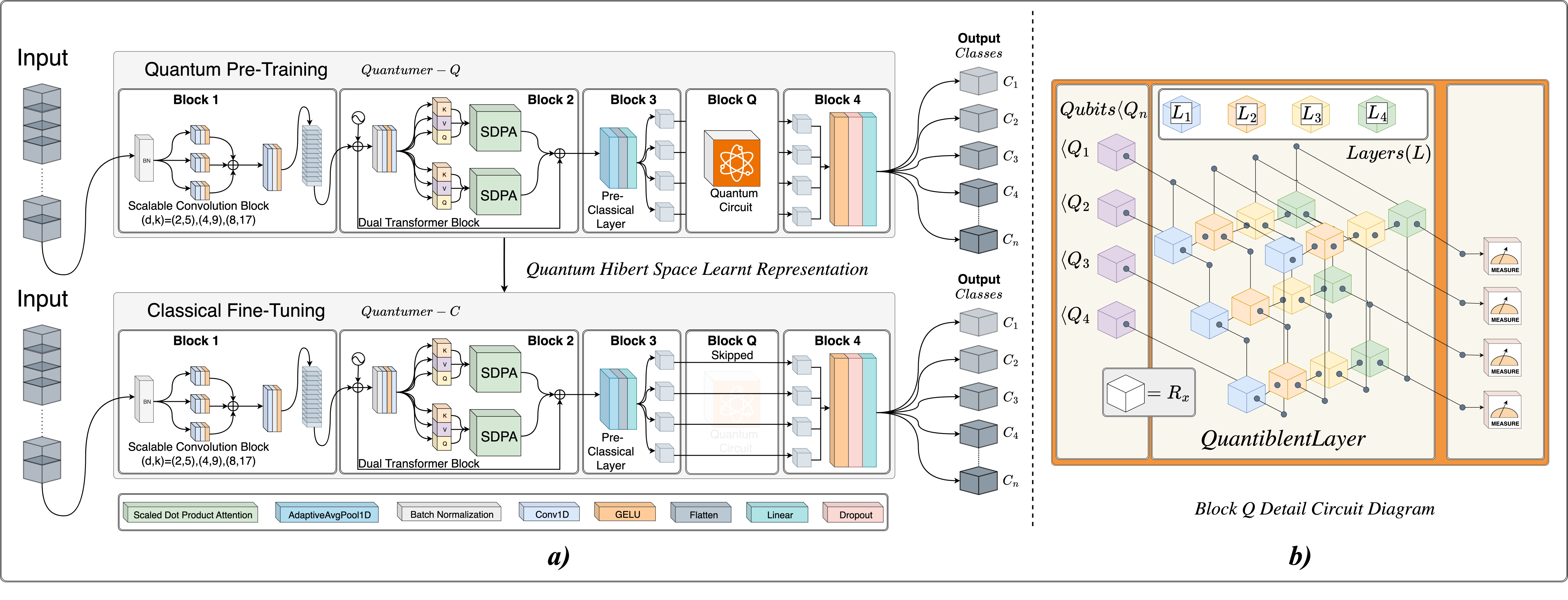}
  \caption{\textbf{Quantumer framework: a two-stage hybrid pipeline for TinyML-based intrusion detection in 6G IIoT networks.} In the quantum pre-training phase (Quantumer-Q), raw sequences undergo multi-scale dilated convolutions and compact Transformer blocks to extract spatio-temporal features, which are pooled and linearly projected into four rotation angles. These angles drive a four-qubit cyclic controlled-rotation variational circuit (Block Q), whose Z-basis measurements yield a compact quantum-assisted feature interface. In the classical fine-tuning phase (Quantumer-C), the same convolutional, Transformer, and pooling stages are retained, but the quantum circuit is bypassed, resulting in a fully classical, resource-efficient TinyML model suitable for 6G edge deployment.}
  \label{Fig:QuantumFramework}
\end{figure*}

\section{Quantum Machine Learning and Hilbert-Space Embeddings}
\label{sec:qml-hilbert}

\subsection{Primer on Variational Quantum Circuits}
Variational quantum circuits (VQCs) are parameterized quantum models in which classical data are encoded into a quantum state, trainable gates transform the state, and measurements produce real-valued outputs. During a forward pass, classical data are encoded using rotation gates, followed by trainable layers that apply entangling operations. Measurements of qubit observables yield expectation values that serve as model outputs~\cite{Biamonte2017}. In Quantumer, the VQC is used as a training-time nonlinear embedding teacher, not as an inference-time accelerator.

\subsection{Quantum versus Classical Feature Maps}
Classical feature maps project data into Euclidean spaces through fixed or trainable nonlinear functions. Quantum nonlinear embeddings (QNEs), in contrast, encode data into quantum states and use measurement statistics as features. Empirical studies show that certain entangling circuits can induce useful representations for supervised learning~\cite{Havlicek2019}. In this work, we do not claim that the deployed model directly uses all quantum amplitudes. Instead, the circuit uses Hilbert-space state evolution during training and exports a compact four-value measurement interface to guide the classical pathway.

\subsection{Hilbert-Space Benefits for TinyML}
Embedding into a quantum Hilbert space can enable nonlinear feature shaping prior to measurement. For resource-constrained devices, this is useful only when the quantum computation does not increase deployment cost. For this reason, Quantumer uses a four-qubit circuit during training and removes it before edge deployment. The four-qubit design provides a small latent quantum state during pre-training and produces four measured outputs that match the lightweight transfer interface. Larger circuits may increase representation capacity, but they also increase training cost and device dependence; therefore, the four-qubit setting is used as a compact training-time feature teacher.

\section{The Quantumer Framework}
\label{sec:quantumer-framework}

In this section, we present Quantumer, a hybrid TinyML--quantum framework for edge-side intrusion detection. As illustrated in Fig.~\ref{Fig:QuantumFramework}, Quantumer follows a two-stage pipeline that combines lightweight TinyML feature extraction with quantum-assisted pre-training. First, packet-level traffic is converted into compact multi-resolution descriptors through a shared classical encoder. These descriptors are then projected into a variational quantum circuit during Quantumer-Q. After pre-training, the quantum block is removed, and Quantumer-C retains only the classical inference path. Thus, the edge device does not execute a quantum circuit and requires no quantum hardware during deployment.

The key design principle is to use the quantum circuit as an offline representation teacher rather than as an inference-time accelerator. This directly supports deployment feasibility: Quantumer-Q may use a quantum simulator or quantum processor during training, whereas Quantumer-C remains a compact classical model for edge inference. As a result, the final inference graph avoids quantum-hardware dependence, circuit-simulation overhead, and runtime measurement cost.

\subsection{Basics of Quantumer: Transfer Learning Overview}

Transfer learning improves a target learner by transferring useful representations from a related source stage. In classical settings, an encoder is commonly pre-trained and then fine-tuned on target data~\cite{Survey_Transfer_Learning_zhuang2020comprehensive}. Quantumer follows this principle, but the source-stage representation is shaped through a compact variational quantum circuit before being transferred to the classical inference model. In Quantumer-Q, the encoder and quantum layer are trained together; in Quantumer-C, the learned interface is retained while the quantum circuit is bypassed. This allows the deployment model to inherit the representation structure learned during quantum-assisted pre-training without increasing inference cost.

\subsection{From Classical Transfer to Quantum Hilbert-Space Learning}

Although classical transfer learning operates in Euclidean feature spaces, Quantumer introduces a quantum-assisted training stage by projecting learned features into a variational quantum circuit. Parameterized rotations and cyclic entangling operations embed the four projected angles into a quantum state, and measurement outputs provide a compact representation for supervised learning. Positive pairs are samples from the same class, while negative pairs are samples from different classes within a mini-batch. During Quantumer-Q, the training process encourages positive pairs to produce closer quantum-state behavior and negative pairs to remain separated, providing a compact interpretation of the fidelity-guided separation mechanism.

This pair-based design is useful for IDS because different attacks can produce partially overlapping flow-level signatures. A model that only fits labels may still place scanning, backdoor, injection, or denial-of-service samples close together in the latent space. Quantumer-Q is intended to reduce this ambiguity by promoting same-class compactness and different-class separation before transferring the representation to Quantumer-C.

The measured quantum output representationaly compact. Although the four-qubit state evolves in a 16-dimensional Hilbert space before measurement, Quantumer does not pass the full statevector to the classifier. Instead, it reads four expectation-based features and uses them as a training signal. This avoids overstating the deployment benefit of Hilbert-space dimensionality and clarifies that the practical contribution lies in training-time nonlinear representation shaping.

\subsection{Significance of Quantumer}

The self-attention mechanism in Transformers can model long-range dependencies, but its computation and memory requirements can be restrictive for TinyML deployment. Increasing model width, depth, othe number of r attention heads may improve representation capacity, but it can also violate edge memory and latency constraints. Quantumer addresses this trade-off by using compact convolutional and Transformer blocks for feature extraction and assigning the quantum circuit only to the training phase. In this way, Quantumer introduces an additional nonlinear representation signal during pre-training without enlarging the deployed classical model.

\subsection{Quantumer: A Two-Stage Transfer Learning Pipeline}

Quantumer consists of two complementary stages. First, Quantumer-Q learns a nonlinear quantum-assisted embedding through a four-qubit variational circuit. Second, Quantumer-C adapts the learned representation using a lightweight classifier head suitable for TinyML deployment on 6G IIoT devices. This two-stage design separates representation learning from inference practicality: Quantumer-Q is trained offline, where higher computational cost is acceptable, while Quantumer-C is deployed under latency, memory, and model-size constraints. The Raspberry Pi 4 evaluation therefore measures the classical INT8 inference path, not a quantum circuit running on the device.

\subsubsection{Encoder Backbone}

Both Quantumer-Q and Quantumer-C share the same encoder backbone, corresponding to Blocks 1--4 in Fig.~\ref{Fig:QuantumFramework}a. The front end uses dilated 1D convolutions with dilation rates 2, 5, 9, and 17 to capture short-, medium-, and longer-range traffic patterns. These multi-resolution features are then processed by compact Transformer blocks with scaled dot-product attention, feed-forward layers, and residual connections to model cross-scale and long-range dependencies~\cite{Vaswani2017}. Global average pooling produces a fixed-length vector, which is projected to interface with the subsequent stages.

The shared encoder ensures that the quantum-to-classical transfer is transparent: the convolutional, attention, pooling, and projection components are retained, while only the quantum circuit is removed during deployment. This prevents Quantumer-C from exceeding the size of the pre-training model and helps isolate the role of the quantum-assisted stage.

\subsection{Quantum Pre-training: Quantumer-Q}
\label{ssec:quantumer-q}

In Quantumer-Q, the pooled sequence representation is mapped to four rotation angles that parameterize a four-qubit cyclic controlled-rotation variational circuit, as shown in the inset of Fig.~\ref{Fig:QuantumFramework}b. Across the circuit layers, data-dependent rotations and cyclic entangling operations shape the representation during pre-training. The final Z-basis measurements yield four expectation values that are used by the classifier during Quantumer-Q. The circuit is optimized using supervised class information so that same-class samples produce closer quantum-state behavior, while different-class samples remain separated by a margin.

The four-qubit choice is deliberate. A larger circuit could increase representational capacity, but it would also increase training cost and reduce practicality for repeated IDS experiments under TinyML constraints. A four-qubit circuit provides a small 16-dimensional latent quantum state before measurement and four measured outputs that match the lightweight transfer interface. This is a practical middle ground: large enough to introduce nonlinear quantum-state evolution during pre-training, but small enough not to dominate the training pipeline.

QuantiblentLayer should therefore be understood as a task-adapted training layer rather than a claim of a new universal ansatz. It uses common variational-circuit components, including rotations, cyclic entanglement, and measurement, but its role is specific: it is attached to the TinyML encoder during Quantumer-Q, trained with class-aware pair behavior, and removed from the inference graph before deployment.

\subsection{Classical Fine-tuning: Quantumer-C}
\label{ssec:quantumer-c}

During classical fine-tuning, the quantum block from Quantumer-Q is removed. The convolutional, Transformer, pooling, projection, and classifier components are retained and fine-tuned as a fully classical model. The retained interface preserves the representation structure learned during pre-training, but no quantum execution is performed during inference. This quantum-to-classical transition allows Quantumer-C to execute only the compact classical pathway, while Quantumer-Q remains an offline representation-learning stage.

This design also explains the accuracy--efficiency behavior observed in the case study. Quantumer-Q generally provides stronger compact-model discrimination because it includes the quantum-assisted training signal. Quantumer-C and its INT8 variants may show modest accuracy reduction after removing the quantum circuit and applying quantization, but they provide the deployment path required for embedded inference. Feature-separability metrics further inspect this transfer by showing whether same-class samples become more compact and different-class samples move farther apart in the learned latent space.

\begin{figure*}[!htbp]
  \centering
  \includegraphics[width=\textwidth]{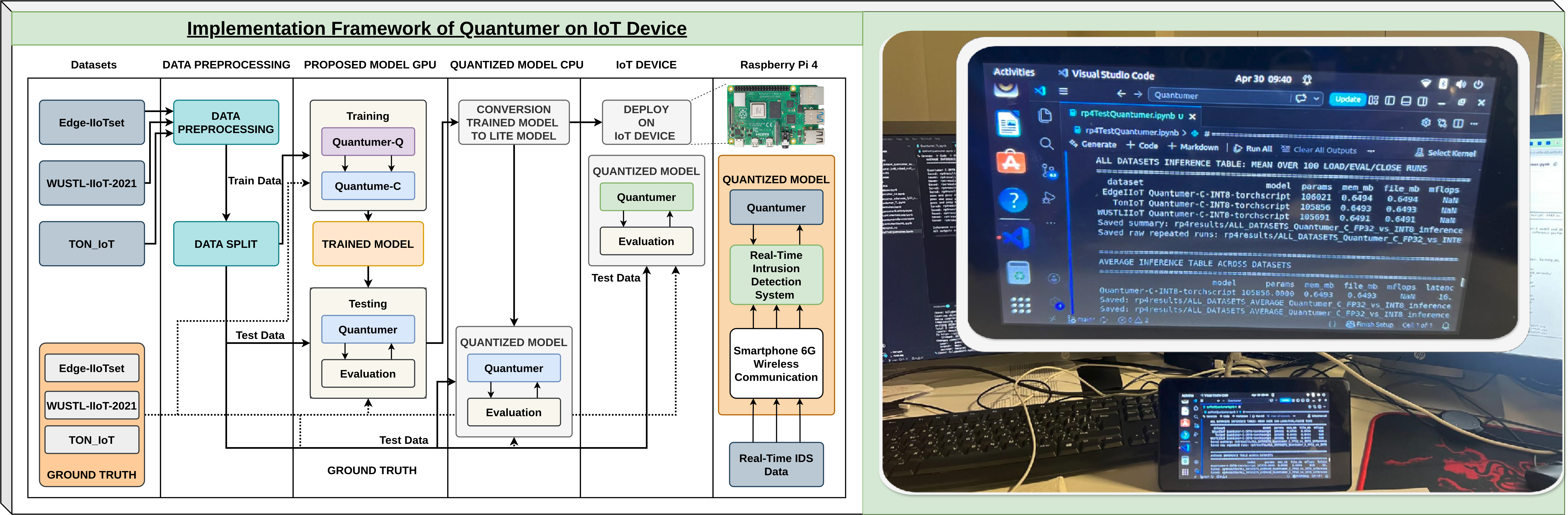}
  \caption{\textbf{End-to-end implementation and deployment workflow of Quantumer on an IoT edge device.} The left panel illustrates dataset preparation, train--test splitting, Quantumer-Q and Quantumer-C training, quantized model conversion, and deployment on Raspberry Pi 4. The right panel shows the real-device setup used to test the deployed INT8 model under practical embedded inference conditions.}
  \label{fig:quantumer_rpi4_pipeline}
\end{figure*}

\begin{figure*}[!htbp]
  \centering
  \includegraphics[width=\textwidth]{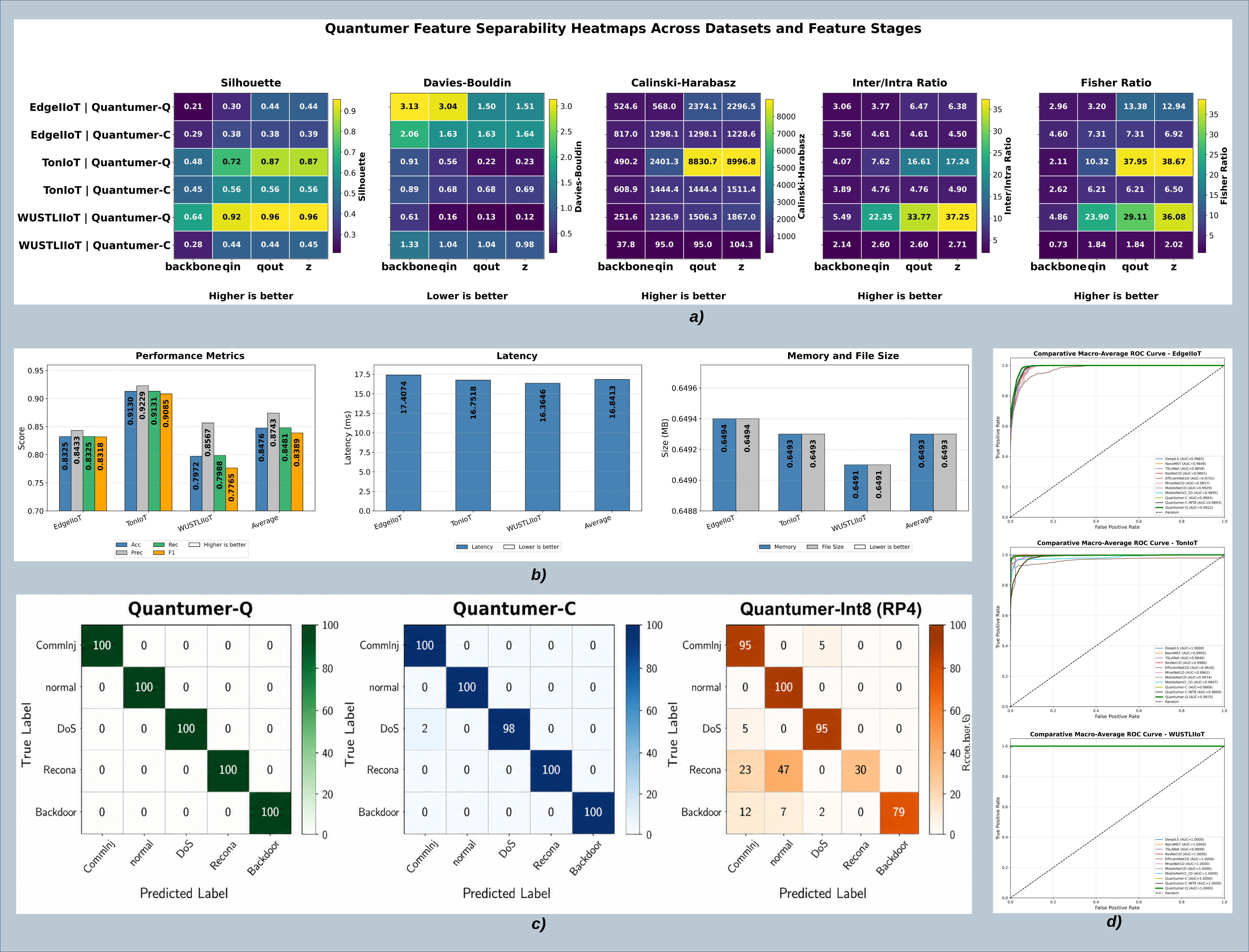}
  \caption{\textbf{Evaluation summary of Quantumer across representation quality, deployment efficiency, and intrusion-detection performance.} (a) Feature-separability heatmaps report Silhouette, Davies--Bouldin, Calinski--Harabasz, inter/intra-class ratio, and Fisher ratio across datasets and feature stages. (b) Deployment-focused comparison of performance, latency, memory usage, model size, and ROC behavior. (c) Confusion matrices compare Quantumer-Q, Quantumer-C, and the INT8 Raspberry Pi 4 deployment. (d) Class-wise ROC curves show the class discrimination ability of the proposed method across Edge-IIoTset, TON IoT, and WUSTL-IIoT-2021.}
  \label{fig:quantumer_results_summary}
\end{figure*}

\section{Case Study: IDS for TinyML-Based 6G IIoT}
\label{sec:case-study}

To validate Quantumer in realistic 6G IIoT security settings, we conduct a case study on three public IIoT/IoT IDS benchmarks. The evaluation focuses on four aspects: dataset relevance to 6G edge environments, feature separability after quantum-assisted pre-training, stability after classical fine-tuning and INT8 conversion, and real-device inference on Raspberry Pi 4. All deployment claims refer to the classical Quantumer-C/INT8 path after the quantum circuit has been removed; Quantumer-Q is used only for offline representation learning.

\begin{table*}[ht]
  \centering
  \caption{Qualitative comparison of feature extraction ability, parameter efficiency, and quantum nonlinear embedding support across classical and proposed models.}
  \label{tab:quantumer_comparison}
  \resizebox{\textwidth}{!}{%
  \begin{tabular}{@{} llccccccc @{}}
    \toprule
    \textbf{Model}
      & \textbf{Feature Extraction}
      & \textbf{Local}
      & \textbf{Global}
      & \textbf{Spatial}
      & \textbf{Temporal}
      & \textbf{Contextual}
      & \textbf{Param.\ Efficiency}
      & \textbf{QNE$^{1}$} \\
    \midrule
    CNN
      & Local convolutions
      & \ding{51} & \ding{55} & \ding{51} & \ding{51} & \ding{55} & \ding{51} & \ding{55} \\
    Transformer~\cite{Vaswani2017}
      & Self-attention
      & \ding{55} & \ding{51} & \ding{55} & \ding{51} & \ding{51} & \ding{55} & \ding{55} \\
    DeepILS~\cite{DeepILS_10873819}
      & Residual channel/spatial attention
      & \ding{51} & \ding{55} & \ding{51} & \ding{51} & \ding{51} & \ding{55} & \ding{55} \\
    TSLANet~\cite{TSLANet_ICML}
      & Adaptive spectral convolution + ICB
      & \ding{51} & \ding{51} & \ding{51} & \ding{51} & \ding{51} & \ding{51} & \ding{55} \\
    NanoMST~\cite{NanoMST_11053971}
      & Multi-scale convolution + Transformer blocks
      & \ding{51} & \ding{51} & \ding{51} & \ding{51} & \ding{51} & \ding{51} & \ding{55} \\
    Quantumer-C (ours)
      & Quantum-enhanced MSA + QNE transfer
      & \ding{51} & \ding{51} & \ding{51} & \ding{51} & \ding{51} & \ding{51} & \ding{51} \\
    \bottomrule
  \end{tabular}%
  }
  \begin{tablenotes}
    \footnotesize
    \item $^{1}$QNE: Quantum Nonlinear Embedding based on Hilbert-space feature representation. For Quantumer-C, QNE denotes training-time transfer from Quantumer-Q and does not imply runtime quantum-circuit execution.
  \end{tablenotes}
\end{table*}

\subsection{Datasets, Preprocessing, and Experimental Setup}
\label{sec:dataset-preproc}

\subsubsection{Datasets}
We evaluate Quantumer on three benchmarks selected to cover heterogeneous sensing, industrial-control traffic, and host-level IoT telemetry. \textbf{Edge-IIoTset}~\cite{Edge-IIoTset_IDS_Dataset} contains smart-city and IIoT traffic with sensor streams, network-flow features, and attacks such as DDoS, spoofing, botnet, and data exfiltration. It reflects dense edge deployments where heterogeneous devices, gateways, and vehicular/IoT nodes generate multimodal telemetry under bandwidth-limited and intermittent-connectivity conditions.

\textbf{WUSTL-IIoT-2021}~\cite{WUSTL-IIOT-2021_IDS_Dataset} provides NetFlow and MODBUS/TCP records from an industrial-control setting, including normal traffic and attacks such as denial-of-service, reconnaissance, command injection, and configuration tampering. It is relevant to ultra-reliable low-latency 6G IIoT scenarios because safety-critical commands may need to be classified at the edge before delayed cloud-side analysis is available.

\textbf{TON IoT}~\cite{TON_IoT_IDS_Dataset} combines network telemetry and host logs from consumer and industrial IoT devices, including DDoS, scanning, ransomware, backdoor, injection, and password attacks. It complements the other datasets by testing whether Quantumer can preserve discriminative representations when evidence is distributed across network and device-level sources. These datasets do not reproduce every physical-layer effect of future 6G systems; rather, they provide IDS benchmarks that capture traffic heterogeneity, attack diversity, edge telemetry, and industrial-control behavior.

\subsubsection{Preprocessing}
All datasets follow the same pipeline: flow-level and host-level feature extraction, normalization of continuous attributes, categorical encoding, temporal windowing, and stratified train--validation--test splitting. The same preprocessing is used for all models so that differences in performance mainly reflect model behavior rather than feature handling. SMOTE is applied only to the training split to reduce class imbalance while avoiding leakage into validation or test data. Window-level aggregation is used because bursty attacks, scans, and short command sequences may not be captured by isolated records, while compact temporal windows remain suitable for TinyML-oriented inference.

\subsubsection{Experimental Setup}
Quantumer is implemented in PyTorch and PennyLane. Training is performed on a GPU server, while deployment validation is conducted on Raspberry Pi 4. In Quantumer-Q, the encoder projects traffic features into a four-qubit variational circuit with four rotation-entanglement layers. The measured representation guides the first training stage. In Quantumer-C, the quantum circuit is removed, and the retained encoder/classifier is fine-tuned as a fully classical model. In this quantum-to-classical transition, the edge device executes only the compact classical pathway and requires no quantum hardware.

For deployment, Quantumer-C is converted into INT8 variants and executed on Raspberry Pi 4, where latency, model size, and memory footprint are measured under embedded CPU conditions, as shown in Fig.~\ref{fig:quantumer_rpi4_pipeline}. We report accuracy, precision, recall, F1-score, ROC-AUC, parameters, memory usage, model size, MFLOPs, and latency. The Raspberry Pi 4 result provides evidence from an embedded CPU for the classical INT8 inference path, while microcontroller-level deployment and quantum-hardware execution remain outside the scope of this evaluation.


\begin{table*}[!t]
\centering
\caption{Comprehensive comparison of Quantumer with baseline models in terms of model complexity, memory footprint, latency, and detection performance.}
\label{tab:main_comparison_quantumer}
\resizebox{\textwidth}{!}{%
\begin{tabular}{lrrrrrrrrrr}
\toprule
\textbf{Model} & \textbf{Params} & \textbf{Mem. MB} & \textbf{Model MB} & \textbf{MFLOPs} & \textbf{Latency ms} & \textbf{Acc.} & \textbf{Prec.} & \textbf{Rec.} & \textbf{F1} & \textbf{ROC-AUC} \\
\midrule
ResNet1D & 4.63M & 5.7412 & 5.8316 & 37.9800 & 0.0248 & \textbf{0.9356} & 0.9327 & \textbf{0.9356} & 0.9292 & 0.9958 \\
\textbf{Quantumer-Q} & \textbf{105.86K} & \textbf{0.4038} & \textbf{0.5525} & 5.5646 & 0.0475 & 0.9331 & 0.9372 & 0.9330 & \textbf{0.9333} & 0.9960 \\
TSLANet & 309.36K & 1.1801 & 1.1897 & 69000 & \textbf{0.0029} & 0.9324 & 0.9364 & 0.9324 & 0.9268 & 0.9961 \\
DeepILS & 2.29M & 5.7412 & 5.8309 & 15.1000 & 0.0238 & 0.9320 & \textbf{0.9400} & 0.9321 & 0.9323 & \textbf{0.9974} \\
NanoMST & 298.00K & 7.2829 & 1.2300 & 7.5900 & 0.0084 & 0.8842 & 0.8997 & 0.8842 & 0.8783 & 0.9917 \\
MobileNetV2\_1D & 2.19M & 8.3660 & 8.5950 & \textbf{5.5012} & 0.0228 & 0.8800 & 0.8938 & 0.8803 & 0.8681 & 0.9947 \\
MnasNet1D & 2.99M & 11.4017 & 11.6417 & 6.9180 & 0.0250 & 0.8686 & 0.8698 & 0.8686 & 0.8536 & 0.9916 \\
MobileNet1D & 3.19M & 12.1563 & 12.2917 & 9.1606 & 0.0155 & 0.8533 & 0.8500 & 0.8532 & 0.8415 & 0.9902 \\
Quantumer-C & 105.86K & \textbf{0.4038} & \textbf{0.5523} & 5.5646 & 0.0086 & 0.8518 & 0.8787 & 0.8523 & 0.8437 & 0.9910 \\
Quantumer-C-INT8-full & 105.86K & 0.5631 & 0.5631 & 5.5646 & 0.0696 & 0.8478 & 0.8755 & 0.8483 & 0.8395 & 0.9906 \\
Quantumer-C-INT8-TorchScript & 105.86K & 0.6493 & 0.6493 & 5.5646 & 0.0829 & 0.8478 & 0.8755 & 0.8483 & 0.8395 & 0.9906 \\
Quantumer-INT8-RPi4 & 105.86K & 0.6493 & 0.6493 & 5.5646 & 16.8413 & 0.8476 & 0.8743 & 0.8481 & 0.8389 & 0.9904 \\
EfficientNet1D & 3.24M & 12.3660 & 12.6214 & 7.7628 & 0.0263 & 0.8408 & 0.8423 & 0.8408 & 0.8180 & 0.9870 \\
\bottomrule
\end{tabular}%
}
\begin{tablenotes}
\footnotesize
\item Note: The Quantumer-INT8-RPi4 row reports real-device embedded CPU deployment on Raspberry Pi 4. Training is performed offline on the GPU server, while deployment uses the classical INT8 inference path after removing the quantum circuit.
\end{tablenotes}
\end{table*}

\subsection{Comparative Models and Baselines}
\label{ssec:baselines}

We benchmark Quantumer against representative lightweight and TinyML-oriented models. DeepILS~\cite{DeepILS_10873819} represents residual convolutional learning with channel/spatial attention. NanoMST~\cite{NanoMST_11053971} is a hardware-aware multi-scale Transformer and serves as the closest compact Transformer baseline. TSLANet~\cite{TSLANet_ICML} represents spectral-convolutional time-series learning. We further include ResNet1D, MobileNetV2-1D, MnasNet1D, MobileNet1D, and EfficientNet1D to compare Quantumer against a broader set of compact one-dimensional convolutional and mobile-efficient backbones.

Table~\ref{tab:quantumer_comparison} provides a qualitative comparison of feature extraction properties across representative classical and proposed models. Classical CNNs mainly capture local spatial/temporal patterns, while Transformers provide global and contextual modeling at higher parameter cost. DeepILS adds channel/spatial attention, TSLANet combines spectral and convolutional modeling, and NanoMST introduces compact multi-scale Transformer blocks. In contrast, Quantumer-C retains local, global, spatial, temporal, and contextual modeling while uniquely incorporating QNE transfer from the quantum-assisted pre-training stage.

Table~\ref{tab:main_comparison_quantumer} then reports the quantitative comparison across model complexity, memory footprint, latency, and IDS performance. Together, these two tables clarify the distinction between Quantumer and classical baselines both conceptually and empirically.

\subsection{Results}
\label{ssec:results}

\subsubsection{Feature Separability}
\label{sssec:feature-separability}

Fig.~\ref{fig:quantumer_results_summary}(a) evaluates representation quality using Silhouette score, Davies--Bouldin index, Calinski--Harabasz score, inter/intra-class ratio, and Fisher ratio. These metrics assess whether latent features form compact same-class clusters and separated different-class clusters. Higher Silhouette, Calinski--Harabasz, inter/intra-class ratio, and Fisher ratio values, together with lower Davies--Bouldin values, indicate better class structure. Because Quantumer-C bypasses the quantum circuit at inference, any retained benefit should be reflected in the training-time representation geometry. The heatmaps show that later feature stages become more structured, suggesting that Quantumer-Q helps organize the latent space before transfer to Quantumer-C.

\subsubsection{Detection Performance}
\label{sssec:detection-performance}

Table~\ref{tab:main_comparison_quantumer} reports the complete comparison of Quantumer with convolutional, mobile-efficient, transformer-based, spectral-convolutional, and deployment-oriented variants. Quantumer-Q achieves competitive compact-model performance with only 105.86K parameters, 0.4038 MB memory usage, 0.5525 MB model size, and 5.5646 MFLOPs. Although ResNet1D obtains the highest accuracy, it requires 4.63M parameters and a larger model footprint. DeepILS achieves the highest ROC-AUC and precision, but it also requires more parameters and memory than Quantumer-Q. TSLANet offers very low GPU-side latency, while NanoMST provides a compact transformer reference but with lower accuracy and F1-score.

The mobile-efficient models, including MobileNetV2-1D, MnasNet1D, MobileNet1D, and EfficientNet1D, provide useful compact backbone references, but their detection scores are lower than those of Quantumer-Q, while requiring substantially more parameters and memory. This indicates that Quantumer-Q occupies a favorable compact-model region: it is not the largest or most computationally expensive model, yet it remains close to the best-performing classical baselines in accuracy, F1-score, and ROC-AUC.

The Quantumer-C and INT8 rows should be interpreted as deployment stages rather than direct replacements for Quantumer-Q. Quantumer-C removes the quantum circuit, thereby trading some accuracy for a fully classical inference path. The INT8-full and INT8-TorchScript variants show the effect of quantization and model conversion, while Quantumer-INT8-RPi4 reports the real-device embedded CPU result. The important observation is that the quantized variants retain high ROC-AUC values while keeping the model footprint below 1 MB, supporting practical edge-side IDS deployment.

\subsubsection{Real-Device Deployment}
\label{sssec:real-device-deployment}

The Raspberry Pi 4 experiment provides real-device validation beyond GPU-only timing. As reported in Table~\ref{tab:main_comparison_quantumer}, Quantumer-INT8-RPi4 keeps the same 105.86K parameter count and a 0.6493 MB model footprint, with 16.8413 ms inference latency on the device. Although slower than GPU-side inference, this measurement is more representative of embedded CPU behavior and supports the feasibility of running the final classical IDS model outside the training server. The result also justifies removing the quantum circuit during deployment: executing a quantum simulator or quantum processor at the edge would be inconsistent with the intended TinyML setting.

\subsubsection{Confusion-Matrix and ROC Analysis}
\label{sssec:confusion-roc-analysis}

Fig.~\ref{fig:quantumer_results_summary}(c) compares the confusion matrices of Quantumer-Q, Quantumer-C, and Quantumer-INT8-RPi4, while Fig.~\ref{fig:quantumer_results_summary}(d) reports class-wise ROC behavior. Quantumer-Q shows the clearest class separation, while Quantumer-C and INT8 retain dominant diagonal behavior with some degradation after quantum-layer removal and quantization. This class-level analysis is important because aggregate accuracy can hide weak minority-class behavior in IDS. The ROC curves further indicate that the quantized variants preserve useful class-ranking information across thresholds, which is relevant when an IDS must tune the operating point to balance missed attacks and false alarms.

\subsection{Discussion}
\label{ssec:discussion}

Quantumer should be interpreted as a two-stage quantum-assisted representation-learning framework rather than a quantum inference accelerator. Quantumer-Q uses the variational quantum layer during offline training to learn compact traffic representations, while Quantumer-C removes the quantum circuit and remains fully classical at deployment, avoiding dependence on quantum hardware and runtime circuit overhead.

Tables~\ref{tab:quantumer_comparison} and~\ref{tab:main_comparison_quantumer} clarify the distinction between Quantumer and prior work. Classical CNNs, Transformers, DeepILS, TSLANet, NanoMST, and mobile-efficient 1D models provide different local, global, spatial, temporal, or contextual modeling capabilities, whereas Quantumer-C uniquely incorporates QNE transfer from the quantum-assisted pre-training stage. Quantumer-Q achieves compact performance with 105.86K parameters and competitive F1-score/ROC-AUC, while Quantumer-C and its INT8 variants provide the deployment path.

The reduction from Quantumer-Q to Quantumer-C/INT8 is expected because the quantum circuit is removed and quantization is applied. However, the retained ROC-AUC values, feature separability results, confusion matrices, and Raspberry Pi 4 measurements indicate that the learned representation remains useful for inference on embedded CPUs. Overall, the results support a focused interpretation of Quantumer as a training-time quantum-assisted representation-learning framework with fully classical deployment, rather than a runtime quantum-acceleration method or a direct full-state quantum inference model.

\section{Conclusion and Future Work}
\label{sec:conclusion}
The Quantumer framework provides a two-stage, quantum-assisted TinyML approach for intrusion detection in 6G IIoT networks. Quantumer-O employs a four-qubit circuit during offline pre-training to learn compact traffic representations. The quantum circuit is subsequently removed, and Quantumer-C performs inference through a fully classical model. The evaluation covers feature-separability heatmaps, confusion matrices, ROC curves, model-complexity analysis, INT8 variants, and Raspberry Pi 4 inference. Quantumer-O uses only 105.86k parameters and requires 5.5466 MFLOPs, whereas the INT8 Raspberry Pi 4 implementation achieves a latency of 16.8413 ms with a memory footprint of 0.6493 MB. These findings support the feasibility of deploying embedded CPUs, but they do not provide evidence of quantum acceleration during inference.

Future work will evaluate larger circuit depths and qubit counts, stricter time- and device-based data partitions, and deployment on lower-power platforms such as TFLite Micro and CMSIS-NN. Future ablation studies will also compare quantum-pre-trained initialization with quantum pre-training without a classical metric or self-supervised objective, as well as random initialization, using the same backbone, preprocessing procedure, and training budget.

\bibliographystyle{unsrt}
\bibliography{references}

\end{document}